\documentclass[onecolumn]{aa} 
\usepackage{graphicx}
\usepackage{txfonts}
\usepackage{natbib}
\bibpunct{(}{)}{;}{a}{}{,}

\begin{document}
   \title{Density structure of an active region and associated moss using Hinode/EIS}

   \author{D. Tripathi\inst{1}, H.~E. Mason\inst{1}, P.~R. Young\inst{2}, G.~Del Zanna\inst{3}}

   \offprints{D. Tripathi@damtp.cam.ac.uk}

   \institute{Department of Applied Maths and Theoretical
              Physics, University of Cambridge, Wilberforce Road,
              Cambridge CB3 0WA, UK \\ 
	      \email{[D.Tripathi; H.E.Mason]@damtp.cam.ac.uk} 
	      \and STFC, Rutherford Appleton Laboratory, Chilton,
              Didcot, Oxfordshire OX11 0QX, UK\\
              \email{P.R.Young@rl.ac.uk} 
	      \and Mullard Space Science Laboratory, University College
              London, Holmbury St. Mary, Dorking, Surrey RH5 6NT, UK\\
	      \email{gdz@mssl.ucl.ac.uk}
} 
  \date{Received 09 Nov 2007; accepted 03 Dec 2007}

 \abstract{Studying the problem of active region heating requires
 precise measurements of physical plasma parameters such as electron
 density, temperature etc. It is also important to understand the
 relationship of coronal structures with the magnetic field. The
 Extreme-ultraviolet Imaging Spectrometer (EIS) aboard Hinode provides
 a rare opportunity to derive electron density simultaneously at
 different temperatures.}{We study the density structure and
 characterise plasma in active regions and associated moss regions. In
 addition we study its relationship to the photospheric magnetic
 field.}{We used data recorded by the EIS, together with magnetic
 field measurements from the Michelson Doppler Imager (MDI) aboard
 SoHO and images recorded with the Transition Region And Coronal
 Explorer (TRACE) and X-Ray Telescope (XRT/Hinode).}{We find that the
 hot core of the active region is densest with values as high as
 10$^{10.5}$~cm$^{-3}$. The electron density estimated in specific
 regions in the active region moss decreases with increasing
 temperature. The moss areas were located primarily on one side of the
 active region, and they map the positive polarity regions almost
 exactly. The density within the moss region was highest at
 $\log\,T=5.8-6.1$, with a value around 10$^{10.0-10.5}$~cm$^{-3}$.
 The moss densities were highest in the strong positive magnetic field
 region. However, there was no such correlation for the negative
 polarity areas, where there was a large sunspot.}{}


\titlerunning{Density structure on active regions}
\authorrunning{D. Tripathi et al.}

\maketitle

\section{Introduction} 

Active regions (ARs) are the brightest features seen on the Sun's surface
when observing in ultra-violet and X-rays. 
Most of the high energy explosions, such as flares and
coronal mass ejections (CMEs), originate from ARs
\citep[e.g.,][]{tripathi04, tripathi06_jaa}. Moreover, studying the
physics of ARs can prove valuable for understanding the
problem of coronal heating. There have been numerous models explaining
coronal heating \citep[see][for a recent review]{klimchuk06}; however,
a solution remains elusive.

From the observational point of view, studying the problem of coronal
heating requires precise and simultaneous measurements of the plasma
parameters such as electron density, plasma flows and non-thermal
broadening etc at different temperatures. Various attempts have been
made to investigate the above-mentioned parameters using previous
instruments such as the Coronal Diagnostic Spectrometer
\citep[CDS;][]{harrison95} aboard the Solar and Heliospheric
Observatory \citep[SoHO;][]{soho}. From diagnostic studies of eclipse
observations and also early X-ray observations it was found that
ARs have a hot and dense core \citep[see e.g.,][]{gabriel,
webb}. \cite{mason99} have produced a density map of an AR from
CDS data using \ion{Si}{x} $\lambda$356.0/$\lambda$374.4 density
diagnostic (formed at $\log\,T=6.1$) and found the highest densities
in the AR core, with values over
2.3$\times$10$^9$~cm$^{-3}$. See other studies with CDS by
\cite{milligan05} and \cite{tripathi06}. It has also been found that the
density and temperatures were higher in the regions of emerging and
cancelling magnetic flux \citep{tripathi06}, similar to the
conclusions derived from early X-ray observations
\citep{webb_zirin}. However, because of limited temperature coverage and
a limited number of density and temperature-sensitive spectral lines,
a comparison of electron densities at different temperatures was not
possible.

\begin{figure}
\centering
\includegraphics[width=0.7\textwidth]{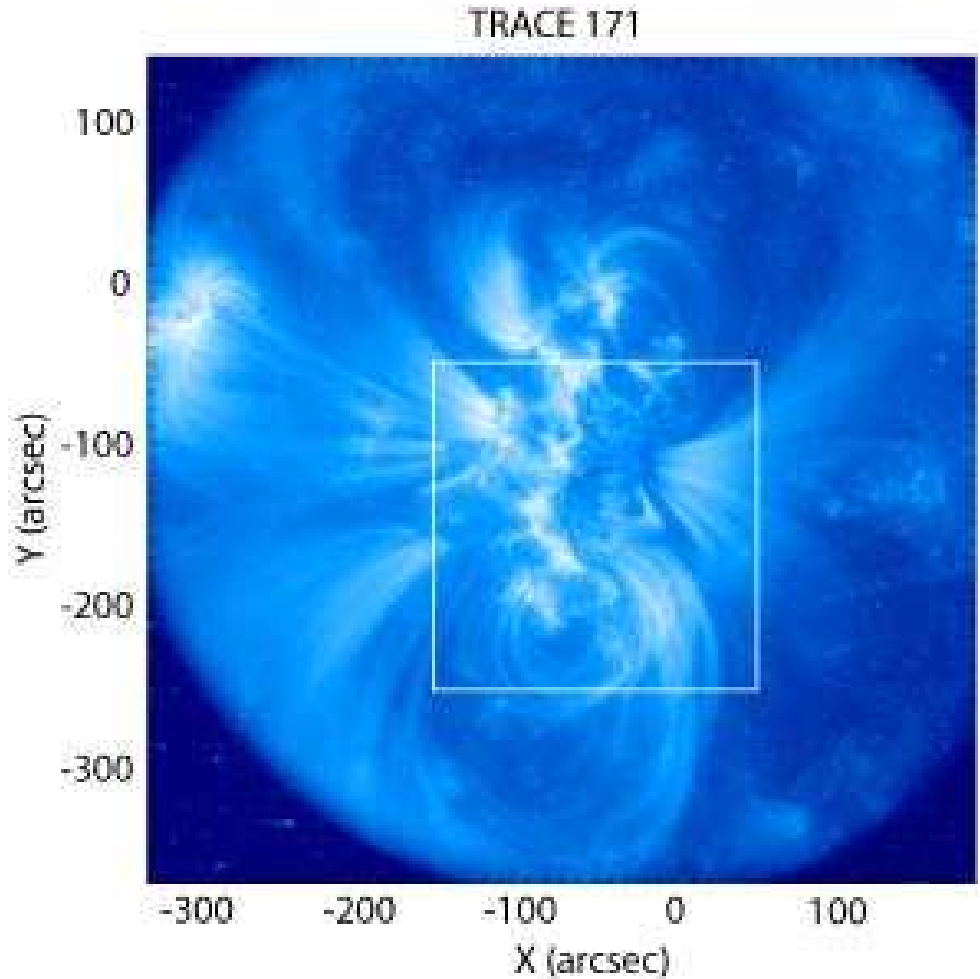}
\caption{TRACE at 171~{AA} showing the AR studied here, 2007 May
  01. The overplotted box shows the region that was rastered by
EIS. \label{complete_ar}}
\end{figure}

\begin{figure*}
\centering
\includegraphics[width=0.75\textwidth]{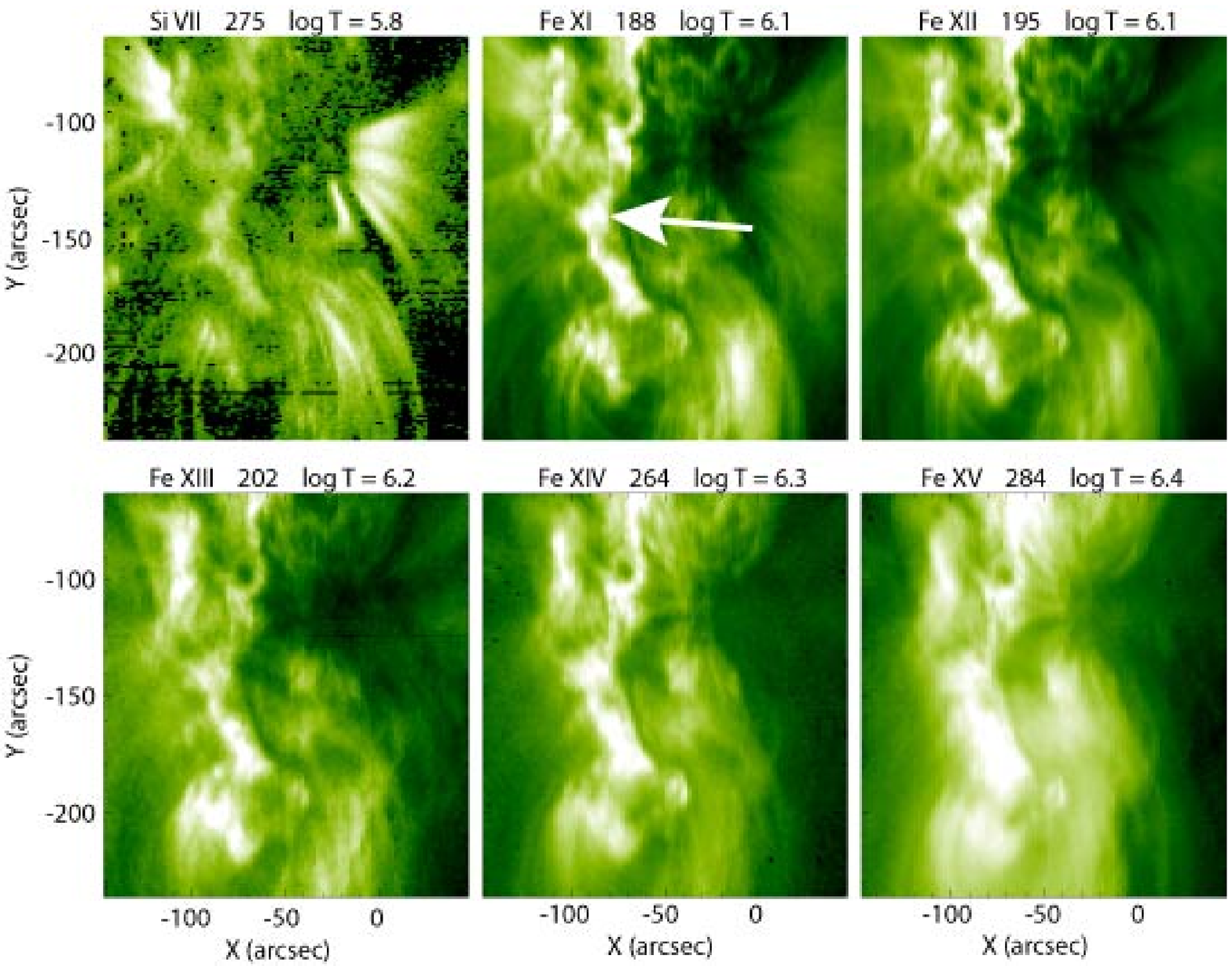}
\caption{Monochromatic images recorded by the EIS. The title of each
panel displays the ion name and the respective wavelengths in which
these images were created and their peak temperature formation. The
arrow in the top middle image indicates a moss region.\label{images}}
\end{figure*}

A specific type of AR emission known as ``moss'' was clearly visible
in 171~{\AA} images (dominated by \ion{Fe}{ix} and \ion{Fe}{x}
emission) taken by the Transition Region And Coronal Explorer
\citep[TRACE;][]{berger_moss, bart_moss}. Morphological and dynamical
aspects of this AR emission are described by \cite{berger_moss}, who
conclude that moss is emission from the upper transition region and
moss regions are the foot points of hot coronal loops \cite[see
also][]{zhao} seen in X-ray emission.  Later, using the spectroscopic
data for an AR from the CDS, \cite{fletcher_moss} show that the moss
region has a temperature range of 0.6$-$1.5$\times$10$^6$~K and is
associated with the footpoints of hot loops. They also show that the
electron density is 2-5$\times$10$^9$~cm$^{-3}$ at a temperature of
about 1.3$\times$10$^6$~K.

In order to characterise the plasma emission from the moss region more
precisely, it is important to simultaneously determine the densities
over a range of temperatures. The Extreme-ultraviolet Imaging
Telescope \citep[EIS;][]{culhane07} aboard Hinode, having very good
spatial and spectral resolution with a broad temperature coverage,
provides an excellent opportunity to study the physical plasma
parameters in ARs, in particular in these moss regions.  Using EIS
data, \cite{warren_07} find that the density in the moss regions can
be as high at 10$^{11}$~cm$^{-3}$ at a temperature of
log~T~$=$~6.1~MK. They did not, however, discuss the association of
the moss with the magnetic field or densities derived over a wide
temperature range.  In this paper we study the density structure of an
on-disk AR and associated moss regions observed on May 1, 2007 using
Hinode/EIS. We determine the density variation as a function of
temperature and also study the relationship between density structure
and photospheric magnetic field structures using data recorded from
the Michelson Doppler Imager \citep[MDI;][]{mdi} aboard SoHO.

\section{Observations and data preparation}
\begin{figure*}
\centering
\includegraphics[width=0.8\textwidth]{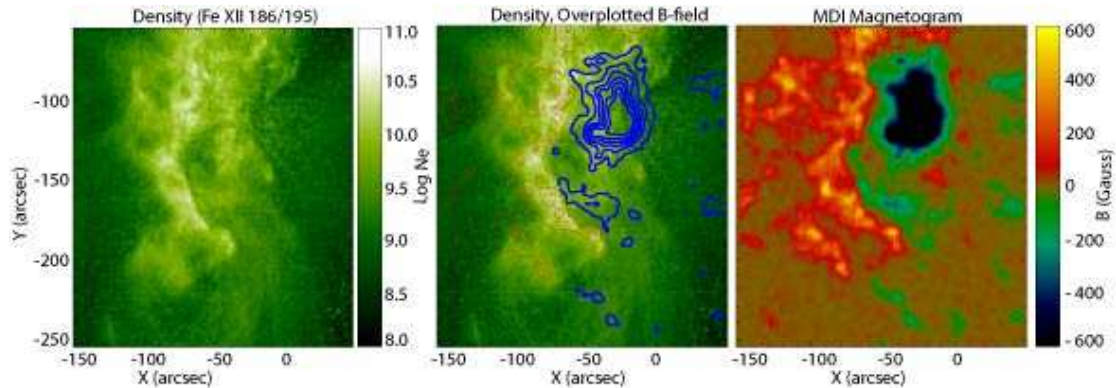}
\caption{Left panel: Electron density map (cm$^{-3}$) derived using
the line ratios \ion{Fe}{xii} 186 and 195~{\AA}. Middle Panel:
Electron density map overplotted with magnetic field contours. Red
contours correspond to positive polarity region whereas blue contours
represent negative polarity regions. Right Panel: MDI magnetogram of
the corresponding region.\label{fe12_dens}}
\end{figure*}

The EIS aboard Hinode provides spectroscopic and imaging observations
of the solar corona and transition region in two wavelength
channels. The first detector covers the wavelength range
246$-$292~{\AA} (CCD-A) and the second covers 170$-$211~{\AA} (CCD-B),
providing observation in a broad range of temperature (log T $=$ 5.8
$-$ 6.7~MK). The EIS provides high cadence images of the transition
region and the corona using 40\arcsec and 266\arcsec
slots. Monochromatic images can be obtained by rastering with a slit
(1\arcsec or 2\arcsec). For other technical details, see
\cite{culhane07}.

In this paper we have used the EIS study we designed,
``cam\_artb\_cds\_a'', which comprises 22 spectral windows covering
spectral lines over a broad range of temperatures. This study uses the
2\arcsec slit with an exposure time of 10~seconds and was run on an
on-disk AR for several days in May 2007. Here we study the
observations made on May 1, 2007. The raster covered a field of view
(FOV) of 200\arcsec$\times$200\arcsec in 20
minutes. Figure~\ref{complete_ar} shows the AR as imaged by the TRACE
171~{\AA} filter. Most of the bright, mottled emission in the core of
the AR (running approximately north/south) comes from areas of
``moss''.  The box in the TRACE image shows the region which was
rastered by EIS using the 2\arcsec slit. The EIS raster mainly covers
the central part of the AR.

The study ``cam\_artb\_cds\_a'' includes many different density
sensitive lines over a range of temperatures. We applied standard
processing routines, namely ``eis\_prep.pro'' and have fitted the
spectrum for each pixel using the routine 'eis\_auto\_fit.pro', both
of which are available in the solar software tree. The line list
published in \cite{tripathi07} was used to derive the electron
density. The electron density values were obtained using the
theoretical line intensity ratios calculated using CHIANTI
\citep{dere97, landi06}.

Some of the spectral lines used in this study are blended with other
lines so care must be taken when deriving the plasma parameters. The
\ion{Mg}{vii}~$\lambda$278 line is blended with a \ion{Si}{vii} line,
the \ion{Fe}{xiii}~$\lambda$203 is blended with another \ion{Fe}{xii}
line and \ion{Fe}{xiv}~$\lambda$274 line is blended with a
\ion{Si}{vii} line. These blends were taken into account when fitting
the lines. The \ion{Fe}{xii} blend from \ion{Fe}{xiii}~$\lambda$203
was removed by fitting a double Gaussian. However, we have used
another \ion{Si}{vii} $\lambda$275, which is quite a strong line, to
remove blends from the \ion{Mg}{vii}~$\lambda$278 and
\ion{Fe}{xiv}~$\lambda$274 lines. The \ion{Fe}{xii}~$\lambda$186 and
$\lambda$195 lines are self blends and care has to be taken when
deriving the densities using \ion{Fe}{xii}~$\lambda$186 and
$\lambda$195 lines, because this can have a substantial influence on high
density regions. For more details on the removal of blends from these
spectral lines, see \cite{peter2}. The \ion{Fe}{xii} line intensities
were problematic for many years, but new work by \cite{zanna} has
resolved these discrepancies. There is an error in one of the
\ion{Fe}{xiii} atomic data files in CHIANTI~v.5.2 that leads to the
\ion{Fe}{xiii} 203/202 ratios yielding incorrect densities. The
corrected file was used for the present analysis and will be made
available in the next CHIANTI release.

For comparing the intensity and density structures with photospheric
magnetic field, we have used magnetograms recorded by the MDI. For
this comparison a co-alignment of the data taken from different
spacecrafts is necessary, which is not straightforward to achieve. In
this analysis for co-aligning purposes, we used the full-disk image
closest in time of observations recorded by the Extreme-ultraviolet
Imaging Telescope \citep[EIT;][]{eit} as a reference. The
EIT~195~{\AA} image was co-aligned with an image of the
\ion{Fe}{xii}~$\lambda$195 line intensities from EIS.  There is a
spatial difference between the images obtained with the two wavelength
bands of EIS (CCD1 and CCD2). The image in the
\ion{Fe}{xii}~$\lambda$195 line emission (CCD1) was co-aligned with
that recorded in \ion{Si}{x}~$\lambda$261 line emission (CCD2).  This
co-alignment was then used for all images from CCD2.  Since images
obtained by MDI are full disk, the pointing information is reliable.

\section{Results and discussion}

\begin{figure}
\centering
\includegraphics[width=0.7\textwidth]{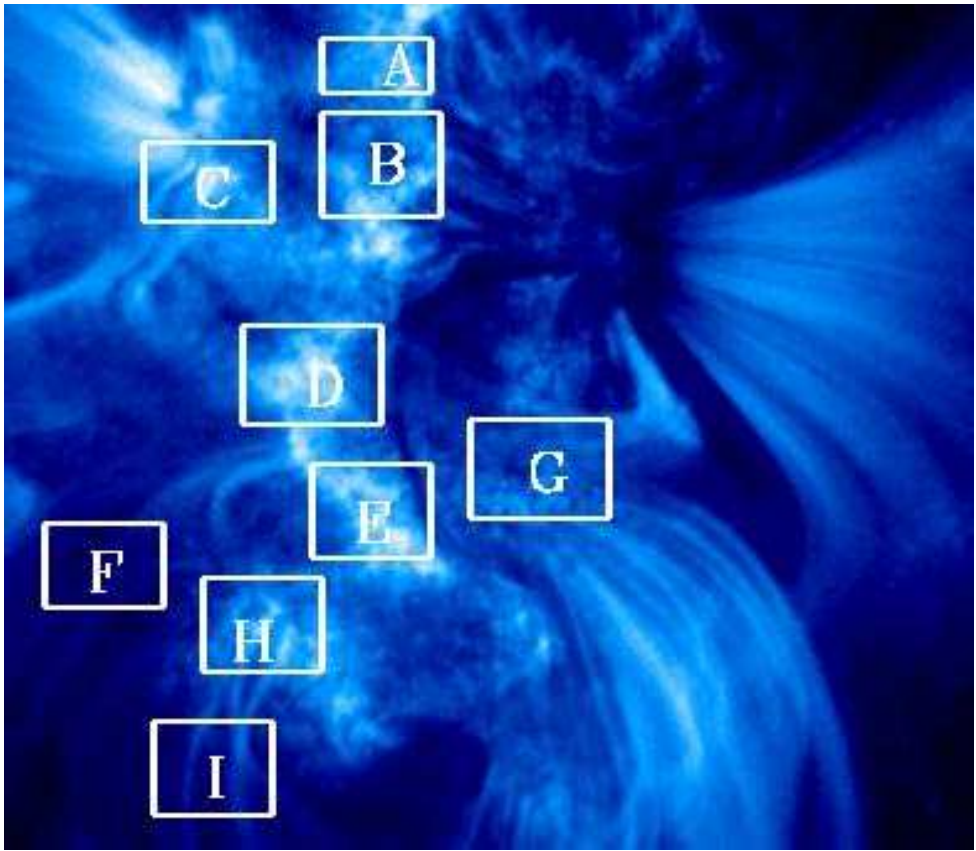}
\caption{TRACE at 171~{\AA} passband showing the regions in which the
average electron densities were investigated further using different line
ratios.\label{trace_reg}}
\end{figure}

Figure~\ref{images} displays monochromatic images recorded
simultaneously using the EIS 2\arcsec slit. The image recorded in the
\ion{Si}{vii}~$\lambda$275 line (top left image) shows the structure
of the AR at the transition region temperature (log~T = 5.8). The very
core of the AR as well as the plume-like structures (on the right-hand
side of the image), which emanate from a sunspot region, can be
clearly seen. The \ion{Si}{vii}~$\lambda$275 line image is very
similar to the corresponding TRACE 171~{\AA} pass-band image shown in
Fig.~\ref{trace_reg}. As we go higher in the temperature such as
\ion{Fe}{xi} (log T = 6.1), \ion{Fe}{xii} (log T = 6.1),
\ion{Fe}{xiii} (log T = 6.2), \ion{Fe}{xiv} (log T = 6.3), and
\ion{Fe}{xv} (log T = 6.4), AR structures become more complex. This
confirms the conclusions from early observations
\citep[e.g.,][]{gabriel, webb, mason99, milligan05, tripathi06} that
the core of an AR is very hot.

The moss emission is seen running north/south in the left-hand side of
the TRACE image (Fig.~\ref{trace_reg}) and also in Fig.~\ref{images}.
It seems that the AR moss is seen clearly only on one side of the
AR. With the EIS (Fig.~\ref{images}), the moss emission can be seen at
many different temperatures, log~T~$=$5.8-6.3~MK.

The left panel of Fig.~\ref{fe12_dens} displays the density map of the
AR derived using the line ratios (\ion{Fe}{xii}, 186.8/195.1). The
\ion{Fe}{xii} line ratios are sensitive to the density range from
10$^7$ to 10$^{12}$~cm$^{-3}$, which provides an opportunity to
measure densities in the core, as well as in the outer regions of
ARs. As is evident from the density map, the density is highest in the
core of the AR and reaches values up to 10$^{10.5}$~cm$^{-3}$.  We
also derived the density maps using line ratios from
\ion{Mg}{vii}~($\lambda$280/$\lambda$278),
\ion{Fe}{xiii}~($\lambda$203/$\lambda$202), and
\ion{Fe}{xiv}~($\lambda$264/$\lambda$274) and found that the densities
in the core of the ARs were highest at all these temperatures
\citep{tripathi07}.

\begin{figure}
\centering
\includegraphics[width=0.7\textwidth]{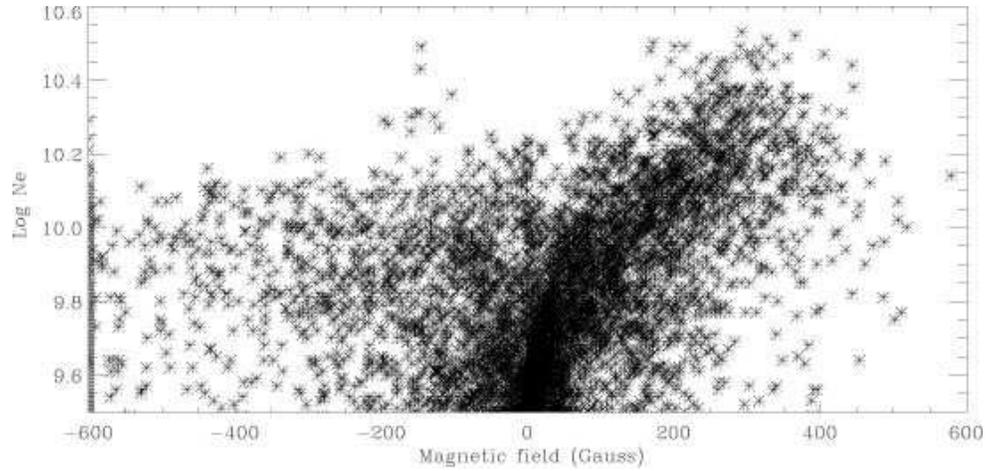}
\caption{Plot showing the variation of electron density (cm$^{-3}$) and
magnetic field strength.\label{mdi_plot}}
\end{figure}

The density maps obtained using line ratios were compared with the
photospheric magnetic field configuration. The middle panel of
Fig.~\ref{fe12_dens} displays magnetic field contours overplotted on
density maps and the right panel displays magnetic field map of the
region. The red contours overplotted on the density map represent the
positive polarity and blue contours represent the negative
polarities. While displaying the magnetic field map of the region, the
magnetic field strength is scaled between $\pm$600~G. We note that in
the positive polarity region, the density maps almost correlate with
the magnetic field strength exactly.  Figure~\ref{mdi_plot} shows a
scatter plot of electron density vs magnetic field strength showing a
strong correlation between electron density and positive magnetic
field strength. There is no similar correlation between electron
density and magnetic field strength for negative polarities. An
interesting point is that the densities are high only towards the
positive polarity side where most of the moss emission is located. On
the negative polarity side, it is possible that the presence of a
large sunspot could have an influence on this phenomenon.


\begin{figure}
\centering
\includegraphics[width=0.8\textwidth]{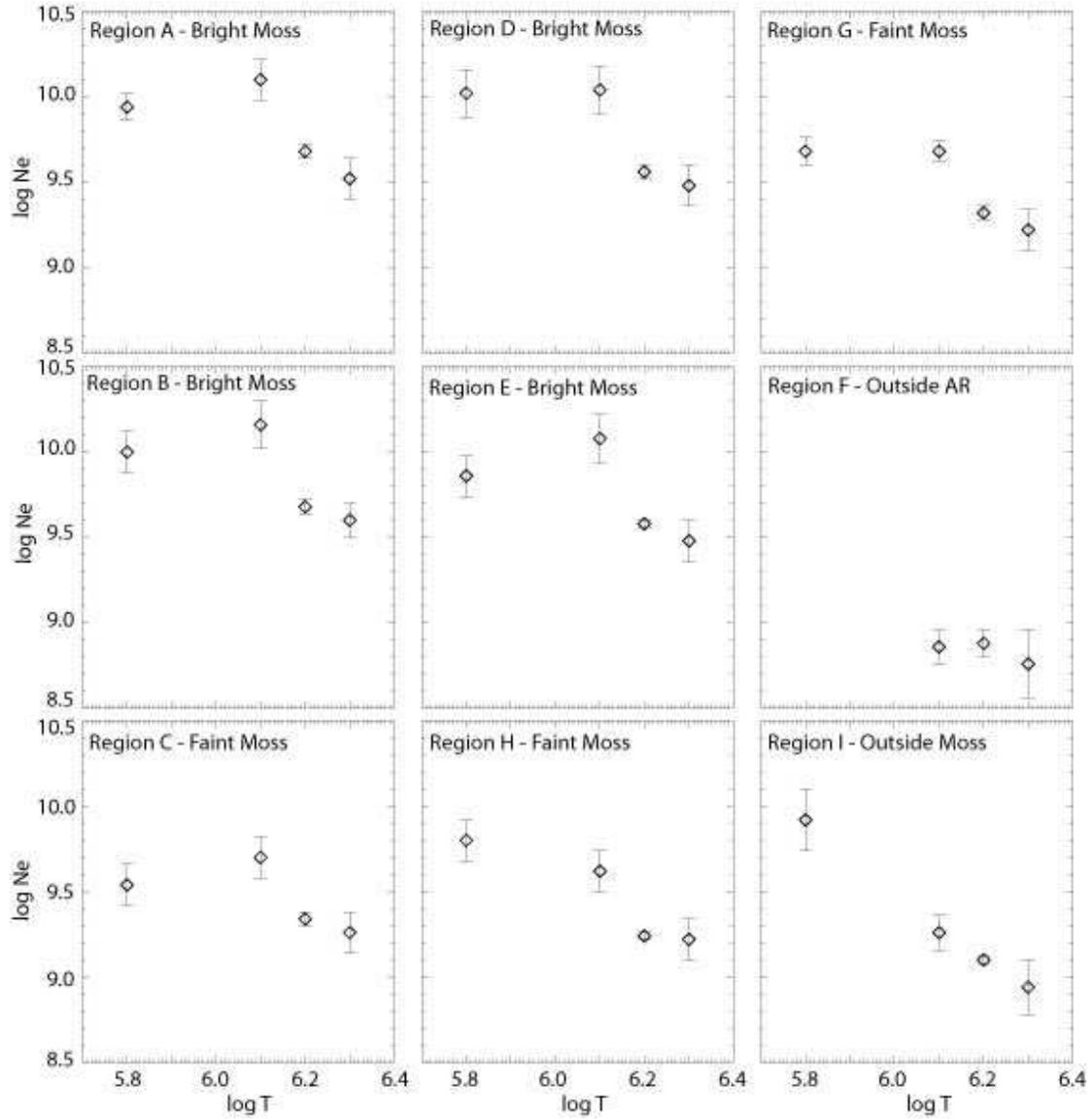}
\caption{Estimated electron densities (cm$^{-3}$) at different temperatures for the regions
marked in Fig.~\ref{trace_reg}. Error bars are estimated by assuming
an error of 10\% in the line intensities.\label{plot}}
\end{figure}

Figure~\ref{trace_reg} displays a TRACE 171~{\AA} image showing the
specific regions inside and outside moss for which we have carried out
a further study of the electron densities using line ratios at
different temperatures. Figure.~\ref{plot} displays average densities
calculated in the marked regions plotted against temperature. Regions
A, B, D, and E are in bright moss regions on the positive polarity
side of the AR. Inside the moss regions, we find that the average
density is about 10$^{10}$~cm$^{-3}$ for log~T$=$5.8~$-$~6.1, and it
drops to around 10$^{9.5}$~cm$^{-3}$ at log~T$=$~6.3. Regions C, H,
and G are in fainter moss regions, which have somewhat lower electron
density values of around 10$^{9.5}$~cm$^{-3}$ at
log~T~$=$~5.8~$-$~6.1. Regions F and I are outside the moss
regions. Coronal loops are evident in region I. As can been seen from
the plots, the electron densities are higher in all regions (inside
and outside moss) at lower temperatures and decrease with
temperature. For region F, we did not have enough counts for
\ion{Mg}{vii} lines and therefore could not derive the densities
at that temperature.  However, the electron density derived in region F
using \ion{Fe}{xii},\ion{Fe}{xiii} and \ion{Fe}{xiv} is just below
10$^{9.0}$~cm$^{-3}$, close to the one for region I at the same
temperatures (log~T$=$6.1-6.3).

\section{Conclusions}

The EIS provides an excellent opportunity to study the physical
plasma parameters simultaneously at many different temperatures from
the transition region to the corona. In this paper we have studied the
density structure in an AR and associated moss observed on
May 1, 2007. We compared the derived densities with the magnetic
field structures observed at the photosphere. We found that the
densities are highest in the core of the ARs across a 
range of temperatures. 

This observations show that the associated moss is located only towards
one side (positive polarity) of the AR.  Using spectral
line ratios, we find that the density inside the moss region is highest
(10$^{10.0-10.5}$~cm$^{-3}$) at log~T~$=$~5.8$-$6.1. The electron density
decreases to 10$^{9.5}$~cm$^{-3}$ at higher temperatures.  In non-moss
regions, where coronal loops are evident, the electron density is
around 10$^{9.0}$~cm$^{-3}$ at log~T~$=$~6.1$-$6.3.  Following a careful
co-alignment, a comparison with the MDI magnetogram reveals that the
high density is correlated with the strong positive field regions
where moss is located. However the negative field region, which
includes a large sunspot, shows no such correlation.

\begin{acknowledgements}
We would like to thank the referee for constructive comments. DT, HEM,
and GDZ acknowledge the STFC. Hinode is a Japanese mission developed
and launched by ISAS/JAXA, collaborating with NAOJ as a domestic
partner and NASA and STFC (UK) as international partners. Scientific
operation of the Hinode mission is conducted by the Hinode science
team organised at ISAS/JAXA. This team mainly consists of scientists
from institutes in the partner countries. Support for the post-launch
operation is provided by JAXA and NAOJ (Japan), STFC (U.K.), NASA,
ESA, and NSC (Norway)
\end{acknowledgements}

\bibliographystyle{aa}
\bibliography{tripathi_ar}
\end{document}